# USE OF UPPER BOUND DELAY ESTIMATE IN STABILITY ANALYSIS AND ROBUST CONTROL COMPENSATION IN NETWORKED CONTROL SYSTEMS

Georges J.-P.[a], Vatanski[b] N., E.Rondeau[a] and S.L. Jämsä-Jounela[b]

[a]) Research Centre for Automatic Control (CRAN UMR 7039) Henri Poincaré University, France
[b]) Laboratory of Process Control and Automation, Helsinki University of Technology, Finland
E-mail: Jean-Philippe.Georges@cran.uhp-nancy.fr  Nikolai.Vatanski@hut.fi  Eric.Rondeau@cran.uhp-nancy.fr

Abstract: Recent interest in networked control systems (NCS) has instigated research in various areas of both communication networks and control. The analysis of NCS has often been performed either from the network, or the control point of view and not many papers exist were the analysis of both is done in the same context. Here a simple overall analysis is presented. In the paper the procedure of obtaining the upper bound delay value in the switched Ethernet network is proposed and the obtained delay estimate is used in stability analysis of the feedback loop and in the control compensation. The upper bound delay algorithm is based on the network calculus theory, the stability analysis uses the small gain theorem, and control compensating strategy is based on Smith predictor, where however the upper bound delay is utilised in obtaining the delay estimate.

## 1. INTRODUCTION

Process automation systems of the future and even those currently in use today, will consist of a large number of intelligent devices and control systems connected by local or global communication networks. In these networked control systems (NCSs) communication between process, controllers, sensors and actuators is performed through the network. The primary benefits from developing the systems from point-to-point systems towards the NCS like systems are reduced system wiring, ease of system diagnosis and maintenance, and increased system agility.

However, for real-time processes, care should be taken when implementing a NCS. In such processes the insertion of the communication network in the feedback control loop introduces an additional either constant or time varying delay that makes the analysis and design of the NCS more complex. Conventional control theories with many ideal assumptions, such as synchronized control and non-delayed sensing and actuation, must be re-evaluated before they can be applied to NCSs.

There are three main directions in approaching the problem of network induced delays in NCS: One way is to design a controller without regard to the delay and then to design a scheduling procedure so that the delay is minimized. The second approach is to study the NCS problem as an integration of network and control design. This paper addresses the third approach where the control strategy is designed so that it compensates a priori the networked-induced delay. During the last years the topic has been actively researched and several compensation strategies have been proposed. Extensive state of art articles and surveys have been published, see (Tipsuwan and Chow, 2003) and (Richard, 2003). The delay compensation methodologies proposed apply ideas from the following control theory fields: robust control (Göktas, 2000), LQG-optimal control (Nilsson, 1998), LMI based control (Li *et at.*, 2004). More specific strategies include: fuzzy logic based control (Almutairi *et al.*, 2001), gain adaptation of controllers (Tipsuwan and Chow, (2002), Smith predictor based compensation (Bauer *et al.*, 2001) to name few.

However, usually in the papers it has been assumed that the information about the network effect on control (delay distribution, uncertainty, deviation from mean value, missing value rate) is known in advance and the information is used in the design or synthesis of the control law. In only few papers the whole procedure of obtaining information about the delay and using it in control system design and synthesis is given. That is the estimation of network properties and using these in control compensation is still usually done in networking and control communities separately.

This paper addresses the gap that still exists between two communities. In this paper the procedure of obtaining information about the delay (the upper bound) in the network is presented and the obtained value is utilised in feedback loop stability analysis and in the control compensation. The delay algorithm presented applies ideas from network calculus theory, the stability analysis uses small-gain theorem and the compensating strategy is based on well known Smith predictor time delay compensation, where however the upper bound delay is used in obtaining the delay estimate based weighting functions from robust control theory.

The paper is organised as follows: Chapter 2 is dedicated to introducing the upper bound delay estimation algorithm. In Chapter 3 stability analysis procedure is presented. In Chapter 4 the delay compensation strategy is introduced. Chapter 5 consists of the results and discussion. The paper ends with a concluding section in Chapter 6.

## 2. UPPER BOUND DELAY ESTIMATION

Here, the switched Ethernet network as an example of NCS network is used. Ethernet network are nowadays more and more used also in control. In this context it is important to understand the behaviour of the network to be able to control the network performances, such as delays (Georges *et al.*, 2004).

Communication network upper bound delay estimation algorithm presented in this paper applies ideas from network calculus theory, see (Cruz (1991); Le Boudec, and Thiran (2001); Jasperneite *et al.*, 2002). For more details of the algorithm, see (Georges *et al.*, 2005).

The communication network is represented as a network of the interconnected switches and each switch is modelled on the other hand as a combination of basic components: multiplexers, demultiplexers and FIFO queues, see Fig. 1.

The traffic coming to the switch, both periodic and aperiodic, is modelled as a "leaky bucket controller". That is, data will arrive at the switch only if the level of the amount of data in the buffer of the switch is less than the maximum buffer size and the data leaves the switch at the constant rate.

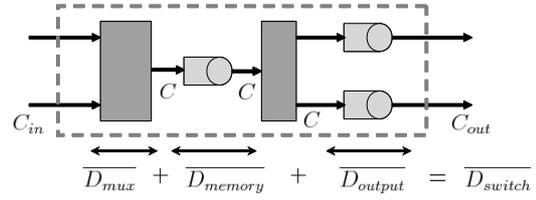

Fig. 1. Model of a 2 ports-switch in full duplex mode based on shared memory and a cut-through management.

Next, the procedure of obtaining the upper bound delay over the network will be explained in more details. In the Section 2.2 it will be explained how to obtain a maximum delay for crossing a single Ethernet switch. And in the following section, the procedure of obtaining end-to-end delays in the network based on the delays over the switches will be given.

*2.1 Maximum delay for crossing Ethernet switch*

To obtain upper bound delay for crossing a single Ethernet switch first the upper bound delays over basic components should be determined. In this Section we will show how to obtain the upper bound delay for FIFO multiplexer, FIFO queue, and demultiplexer basic components. The upper bound delay over the switch is then the sum of the upper bound delays over the basic components:

$$\overline{D_{switch}} = \overline{D_{mux}} + \overline{D_{queue}} + \overline{D_{output}} \quad (1)$$

*Upper bound delay over a FIFO multiplexer.* The first step in calculating the delay over a multiplexer is determining the arrival curves of the traffic coming to the component and the service curves provided by the component. With the assumption that the traffic follows the leaky bucket controller, these curves are affine and have form of:

$$b(t) = \sigma + \rho t \quad (2)$$

Where $\sigma$ is maximum amount of data that can arrive in a burst and $\rho$ is an upper bound of average rate of the traffic flow. The typical arrival and service curves are shown in Fig. 2.

The next step is determining the upper bound backlog in the multiplexer. The backlog is the number of bits accumulated in the component and it is a measure of congestion over the component. For the arrival and service curves in Fig. 2 the upper bound backlog occurs at time $t$ and can be calculated from

$$b_1(t) + b_2(t + L/C_2) - C_{out}t \quad (3)$$

Where $b_1$ and $b_2$ are arrival curves of stream 1 and 2 at time $t$, $L$ is the maximum length of frames, $C_2$ is capacity of import port 2, and $C_{out}$ is capacity of the output link.

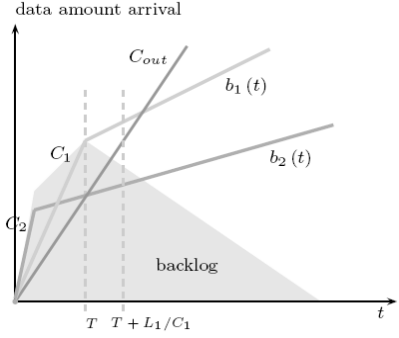

Fig. 2. Arrival and service curves and backlog evolution inside two-input FIFO multiplexer.

When the upper bound backlog over the component is known, the upper bound delay over the multiplexer component is then obtained by dividing the maximum backlog value by the capacity of the output link of the multiplexer.

This procedure can be summarized as follows:

In a FIFO m-inputs multiplexer the delay for any incoming bit from the stream *i* is upper-bounded by:

$$\overline{D_{mux,i}} = \frac{1}{C_{out}} \min_k \overline{B_{mux,k}} \qquad (4)$$

Where $\overline{B_{mux,k}}$ is upper-bound of the backlog in the bursty periods $u_k$.

For $k = i$, (that is $b_i$ is bigger than $b_k$) the bursty period is defined by $u_i = \sigma_i/(C_i - \rho_i)$ and the backlog is upper-bounded by :

$$\overline{B_{mux,i}} = \sum_{z=1; z\neq i}^{m}\left(\sigma_z + \rho_z\left(u_i + \frac{L_z}{C_z}\right)\right) + u_i(C_i - C_{out}) \qquad (5)$$

Where $\sigma_i$ is burstiness of stream *i*, $\rho_i$ is average rate of arrival of data of stream *i*, $L_i$ is maximum length of frames of stream *i* and $C_i$ is capacity of import port *i*.

For $k \neq i$ (that is $b_i$ is smaller than $b_k$) such $1 \leq k \leq m$, we have $u_k = \sigma_k/(C_k - \rho_k) - L_k/C_k$ and

$$\overline{B_{mux,i}} = \sum_{z=1; z\neq k}^{m}\left(\sigma_z + \rho_z\left(u_k + \frac{L_z}{C_z}\right)\right) + u_k(C_k - C_{out}) \\ - \rho_i\frac{L_i}{C_i} + L_k \qquad (6)$$

*Upper bound delay over a FIFO queue.* For the FIFO queue the delay of any byte is upper-bounded by:

$$\overline{D_{queue}} = \frac{1}{C_{out}}\frac{(C_{in} - C_{out})}{C_{in} - \rho_{in}}\sigma_{in} \qquad (6)$$

*Upper bound delay over a demultiplexer.* The demultiplexer has one input link and two or more output links. Its function is to split the streams that arrive to the input ports and to route them to the appropriate output ports. In Ethernet, this consists of reading the destination address at the start of the frame and to selecting the output port associated to its destination in the forwarding table. Due to the Spanning Tree Protocol, only one path is activated to go from one point to another. Therefore it is assumed that the routing step is instantaneously achieved. Thus the demultiplexer does not generate delays.

## 2.2. Maximum end-to-end delays for crossing a switched Ethernet network

In the previous section, upper-bounded delay equations for crossing a switch have been proposed. In the equations, the maximum delay value $\overline{D}$ depends on the leaky bucket parameters: the maximum amount of traffic $\sigma$ that can arrive in a burst and the upper bound of average rate of the traffic flow $\rho$. In order to calculate the maximum delay over the network, it is necessary that the envelop $(\sigma, \rho)$ is known at every point in the network. However, as shown in the Fig. 3, only the initial arrival curve values $(\sigma^0, \rho^0)$ are known and the values for other arrival curves should be determined. To calculate all arrival curve values the following equations can be used:

$$\begin{aligned}\sigma_{out} &= \sigma_{in} + \rho_{in}D \\ \rho_{out} &= \rho_{in}\end{aligned} \qquad (7)$$

For example for the arrival curve $(\sigma^1, \rho^1)$ in Fig. 3 the envelop after the first switch is:

$$(\sigma^1, \rho^1) = (\sigma^0 + \rho^0 \overline{D_{switch}}, \rho^0) \qquad (8)$$

Now it possible to summarize the procedure of obtaining the maximum end-to-end delays in a switched Ethernet network. The algorithm is the following:

1. Identify all streams on each station and determine the initial leaky bucket values.
2. Identify the route of each stream. In switched Ethernet networks, paths are determined by the spanning tree protocol.
3. On each switch, formulate output burstiness equations for all streams.
4. Define the equation systems of form $A\Psi = \Phi$ or $a_n\sigma_1 + b_n\sigma_2 + ... + z_n\sigma_m = \delta_n$
5. Solve the burstiness values.
6. Determine the end-to-end delay in the network from the equation

$$\overline{D_i} = \frac{\sigma_i^h - \sigma_i^o}{\rho_i} \qquad (9)$$

where *h* is the number of crossed switch.

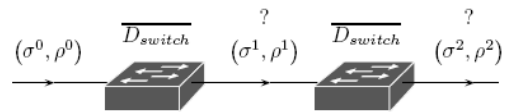

Fig. 3. Burstiness along a switched Ethernet network.

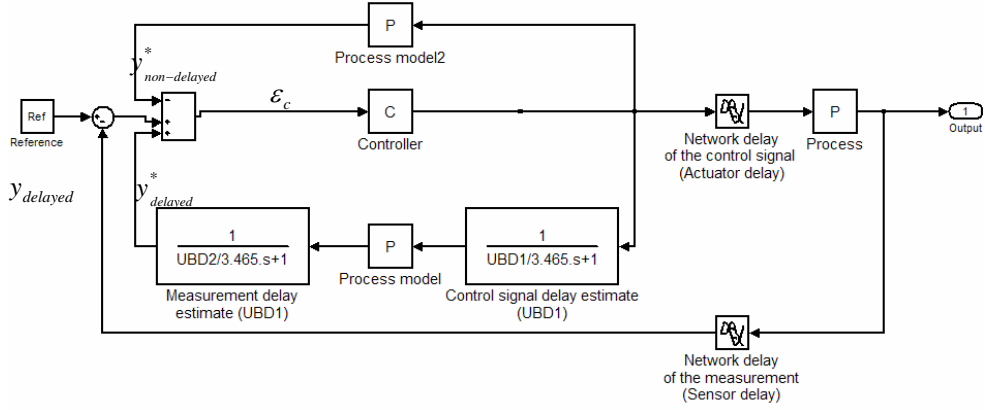

Fig. 4. A Smith predictor for compensation of the network induced delay

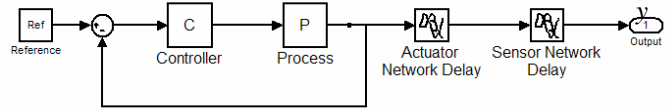

Fig. 5. The equivalent Smith predictor compensation scheme in case the model and the time delays are assumed known

## 3. STABILITY ANALYSIS USING UPPER BOUND DELAY ESTIMATE

The upper bound delay estimate obtained can be utilised in checking whether the feedback system may become unstable under such delay. In this section a simple procedure of stability analysis of control loop with time varying delays is shortly presented. The criterion used is from (Kao and Lincoln, 2004).

Consider the single-input–single-output system with plant $P$ and controller $C$, and the time varying delay in the control system. (The delay can be placed anywhere in the loop)

Assuming that the closed loop system is stable for zero delay, the following theorem gives simple criteria of stability for the system with an arbitrarily time-varying but bounded delay. For the closed loop system with continuous-time $P(s)$ and $C(s)$, the system is stable if the following inequality holds:

$$\frac{P(jw)C(jw)}{1+P(jw)C(jw)} < \frac{1}{UBD \cdot w}, \forall w \in [0,\infty] \quad (10)$$

Where $P$ and $C$ transfer functions for the process and controller without delays, and $UBD$ is upper bound delay estimate.

That is the frequency response of complementary sensitivity function ($T$) is calculated and compared with the stability limit. If at some point the $T$ is bigger than the limit, the closed loop system may become unstable at that frequency. In Bode diagram this can be seen as an interception between $T$ and the stability limit curves.

## 4. DELAY COMPENSATION USING UPPER BOUND DELAY ESTIMATE

In the NCS environment the main goal of the control system is to maintain the control and system performance as much as possible regardless of the delays in the network. That is the system should robust to the delay induced by the network.

The control compensation strategy used in this paper is based on well known Smith predictor. The compensation scheme is shown in Fig. 4. In the figure minor feedback loops have been introduced around the conventional controller.

To see, how the scheme works, let us proceed to analyze the time-delay compensator assuming that there are no model errors in the scheme (process model and time delay are known exactly). This gives that the delayed process measurement and the delayed process output estimate are equivalent, $y_{delayed} = y^*_{delayed}$. Then observe that the signal reaching the controller is a corrected error signal given by:

$$\varepsilon_c = r - y_{delayed} + y^*_{delayed} - y^*_{non-delayed} \quad (11)$$

$$\text{or } \varepsilon_c = r - y^*_{non-delayed} \quad (12)$$

That is the error signal that reaches the controller is calculated based on the non-delayed estimate of the process output. Implying, as a result, that block diagram of Fig. 4 is equivalent to that shown in Fig. 5. The net result of introduction of minor loop is therefore to eliminate the time delay factor from feedback loop where it causes stability problems and move it outside the loop, where it has no effect on closed-loop stability.

The scheme works well as long as the process model and time delay are known. In case there are modeling errors, they affect the performance of Smith predictor. In addition the Smith predictor scheme is designed for constant time delays and therefore may not perform as well for systems with time delays which significantly vary over time.

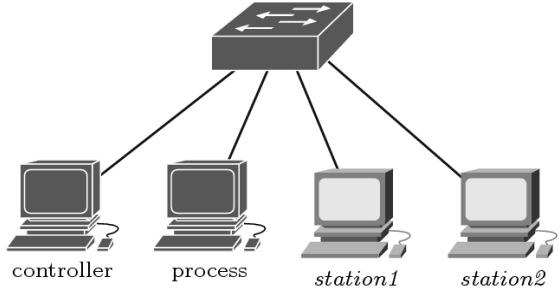

Fig. 6. The structure of the network

To increase the robustness of the Smith predictor, so that it is more suitable for the NCS environment were time delay varies, we use the following delay estimate obtained based on upper-bound delay value. This delay estimate has been originally proposed by Wang, *et al*., (1994) to represent uncertain delays in $H_\infty$ framework in the design of robust controllers.

The delay can be approximated by

$$e^{-\tau \Delta s} \approx 1 - \frac{\tau s}{1+\frac{\tau}{2}s}\Delta = 1 + w(s)\Delta, |\Delta| \leq 1 \quad (13)$$

Where $\Delta$ describes uncertainty. However, instead of using $w(s)$ directly Wang, *et al.* proposed to use $w_h$ which is more robust to the delay errors:

$$w_h(s) = \frac{UBD \cdot s}{1 + UBD \cdot s/3.465} \quad (14)$$

## 5. TESTING RESULTS AND DISCUSSIONS

In the testing the network of four stations that were connected over a full duplex Ethernet switch was used. One station was dedicated for the real time process, one for the controller and two for generating overload traffic in the network. The structure of the system is shown in Fig. 6.

To calculate the upper bound delay, first, the initial leaky bucket values of each stream were identified. There are 6 messages sent periodically. The traffic sent from the process to the controller is given by $b_1^0(t)$, the traffic from the controller to the process by $b_2^0(t)$. These traffics correspond to the frames of 72 bytes sent every 10ms. Upper-bounds for these traffics will be computed to obtain upper bounds, $UBD_1$ and $UBD_2$. We consider also background traffic ($b_3^0(t)$, $b_4^0(t)$, $b_5^0(t)$, $b_6^0(t)$) from the stations to the process and to the controller in order to overload the network:

$$b_1^0(t) = b_2^0(t) = \sigma_1^0 + \rho_1 t = 72 + 7200t$$
$$b_3^0 = b_4^0(t) = b_5^0(t) = b_6^0(t) = \sigma_3^0 + \rho_3 t = 1526 + 305200t$$
(15)

Next, the route of each stream was identified and output burstiness equations were formulated. After solving the burstiness values the end-to-end upper bound delay for streams 1 and 2 are:

$$UBD_1 = UBD_2 = \frac{\sigma_1^2 - \sigma_1^0}{\rho_1} \approx 3.5\,ms \quad (16)$$

Next, the upper bound delay was used in checking whether the system can become unstable under such delay. Consider the following real-time process and controller (time in *ms*):

$$P(s) = \frac{2}{(s+5)(s+0.2)} \quad (17)$$

$$C(s) = \frac{K_P s + K_I}{s}, \quad K_P = 0.5, \quad K_I = 0.5$$

Using the upper bound delay the stability criteria presented in Eq. (10) for was calculated. The Bode diagram of the closed loop transfer function and the stability limit are shown in Fig. 7. From the figure it can be seen that the frequency response curve of the closed loop transfer function and the stability limit curve intercept. That is, at frequencies between 0.2-9 rad/ms the stability criteria doesn't hold, and the feedback control loop becomes unstable at this frequency range. Thus, a delay compensation strategy is needed.

The delay compensation strategy based on Smith predictor presented in Fig. 4 was used. The model was assumed to be known, and the network delay in sensor and in actuator sides was assumed to vary randomly between zero and upper delay value estimate. The simulation result is presented in Fig. 8. In the figure three signals are shown; the set point, the system output in case no delay compensation is used, and the system output with the compensation strategy implemented. It can be concluded that even with such simple delay compensation strategy the effect of the network induced delay on the feedback control loop can be significantly reduced.

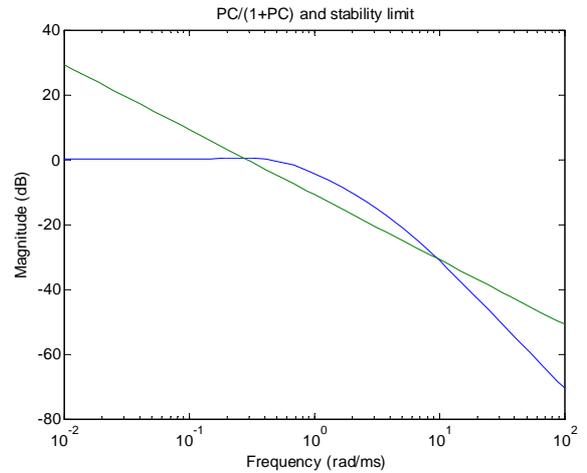

Fig. 7 Stability checking for the control system and upper bound delay estimate using the criteria given in Eq. (10)

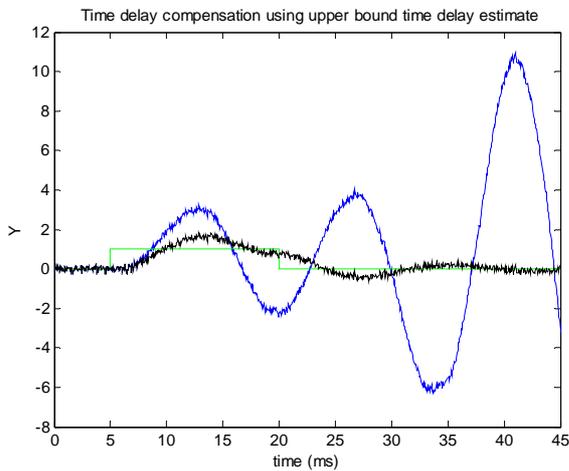

Fig. 8. System output, when there is no delay compensation and with delay compensation

## 6. CONCLUSION

In this paper a simple overall analysis of the networked control system has been presented. A procedure of obtaining the upper bound delay value in the switched Ethernet network is presented and the obtained delay estimate is used in stability analysis of the feedback loop and in the control compensation. It can be concluded that upper bound delay estimate is an important measure of networked control system that can also be used design and synthesis of the control system.

## ACKNOWLEDGEMENTS

This research has been conducted as a part of the Networked Control Systems Tolerant to Faults (NeCST) project IST-004303 that is partially funded by the EU. The authors gratefully acknowledge the support.